\documentclass[showpacs,twocolumn]{revtex4}

\usepackage{graphicx}
\usepackage{dcolumn}
\usepackage{bm}
\usepackage{amsfonts}
\usepackage{amssymb,epsf}

\begin{document}

\title{Nonstatic charged BTZ-like black holes in $N+1$ dimensions}
\author{Sushant~G.~Ghosh}\email{sghosh2@jmi.ac.in, sgghosh@gmail.com}

\affiliation{Centre for Theoretical Physics, Jamia Millia Islamia,
New Delhi - 110025, INDIA}

\begin{abstract}
We find an exact nonstatic charged BTZ-like solutions, in
($N$+1)-dimensional Einstein gravity in the presence of negative
cosmological constant and a nonlinear Maxwell field  defined by a
power $s$ of the Maxwell invariant, which describes the
gravitational collapse of charged null fluid in an anti-de Sitter
background. Considering the situation that a charged null fluid
injects into the initially an anti-de Sitter spacetime, we show that
a black hole form rather than a naked singularity, irrespective of
spacetime dimensions, from gravitational collapse in accordance with
cosmic censorship conjecture. The structure and locations of the
apparent horizons of the black holes are also determined. It is
interesting to see that, in the static limit and when $N=2$, one can
retrieve $2+1$ BTZ black hole solutions.
\end{abstract}

\pacs{04.40.Nr, 04.20.Jb, 04.70.Bw, 04.70.Dy}

\maketitle

\section{introduction}
Interest in 2 + 1 dimensional gravity has been intensify by the
discovery of a black hole solution by Banados \textit{et al.}
\cite{BTZ1,BTZ2} (referred as BTZ black hole). It enjoys many
properties of its counterparts in 3+1 which makes BTZ a suitable
model to understand black hole physics in a quite simpler setting.
The BTZ black hole is a solution of the $(2+1)$-dimensional
Einstein-Maxwell gravity with a negative cosmological constant
$\Lambda =-1/\chi^{2}$. The metric is given by \cite{BTZ1,BTZ2}:
\begin{equation}
ds^{2}=-N^{2}(r)dt^{2}+\frac{dr^{2}}{N^{2}(r)}+r^{2}d\varphi ^{2},
\label{BTZmetric}
\end{equation}
with  \[N^{2}(r)=\frac{r^{2}}{\chi^{2}}-\left[ M+2q^{2}\ln
(\frac{r}{\chi})\right],\]  where $N^{2}(r)$\ is known as the lapse
function and $M$ and $q$ are the mass and electric charge of the BTZ
black hole, respectively.

The nonlinear electrodynamics models have been proved to be
excellent testing grounds in order to elude some problems that occur
in the standard Maxwell theory. Indeed, the interest for nonlinear
electrodynamics started with the  work of Born and Infeld \cite{bi}.
There has been significant interest about black hole solutions
\cite{mhcm,mhcm1,mhcm2,hendi,hendi2,hendi3,hendi4,hendi5,hendi6,
Breton,Hendi1}, including regular black holes \cite{ag,ag1}, with
nonlinear electrodynamics source given as an arbitrary power $s$ of
the Maxwell invariant, i.e., $\left(F_{a b }F^{a b }\right) ^{s}$.
The nonlinear electrodynamics coupled to general relativity are
explored and it has been derived black-hole solutions with
interesting asymptotic behaviors \cite{mhcm,mhcm1,mhcm2}. Hendi
\cite{Hendi1} demonstrated that Einstein-nonlinear Maxwell gravity
can generates BTZ-like solutions in $(N+1)$-dimensional solutions.
He showed that the electric field of BTZ-like solutions is the same
as $(2+1)$-dimensional BTZ black holes, and also their lapse
functions are approximately the same.

Over the past decade there has been an increasing interest in the
study of black holes, and related objects, in higher (and lower)
dimensions, motivated to a large extent by developments in string
theory. The aim of the present paper is to obtain nonstatic
higher-dimensional analogue of well known BTZ black hole solution in
2 + 1 dimensions.  To be precise, we are interested in
$(N+1)$-dimensional exact nonstatic Vaidya-like solution of the
Einstein gravity coupled with nonlinear Maxwell source in the
presence of null fluid thereby by generalizing those discussions in
\cite{Hendi1}.  The Vaidya geometry permitting the incorporations of
the effects of null fluid or null dust offers a more realistic
background than static geometries, where all back reaction is
ignored.  One of  the important null fluid solutions include the
Robinson - Trautman \cite{rt} solution which include, as a special
case, the Vaidya solution \cite{pc,pc1} and in turn include the
Schwarzschild vacuum solution.
 Also, several extension of Vaidya \cite{pc,pc1}
solution, in which the source is a mixture of a perfect fluid and
null fluid, have been obtained in later years both in four
dimensions \cite{ww,adsg,ka} and higher dimensions \cite{sgad,sgh}.
This includes the Bonnor-Vaidya solution \cite{bv} for the charge
case. The Vaidya solution \cite{pc} are widely used as a testing
ground for various gravitational scenario and formulation of the
cosmic censorship conjecture (CCC) \cite{psj,psj1}.  The CCC
proposes that singularities are always hidden within event horizon,
and therefore cannot be seen from the rest of spacetime, i.e., no
naked singularities. For the ultimate fate of gravitational collapse
we could still do no better than the  CCC \cite{rp,rp1,psj,psj1,rm}.
The CCC, put forward by Penrose 40 years ago, is still one of the
most important open questions in general relativity.  So far the
weak CCC has not been seriously challenged while there exist many
counter examples challenging the stronger version  of CCC \cite{rm}.

There are several issues that motivate our analysis: how does the
nonlinear Maxwell source change the final fate of collapse?  Whether
such solutions lead to naked singularities? Do they get covered due
to departure from spherically symmetry? Does the nature of the
singularity changes in a more fundamental theory preserving cosmic
censorship? As we will see, these collapsing solutions do have
several interesting features.
\section{Field Equations and their solutions} The
$(N+1)$-dimensional ($N\geq2$) action in which gravity is coupled to
nonlinear electrodynamics field reads \cite{mhcm,mhcm1,mhcm2}
\begin{equation}
\mathcal{I}=-\frac{1}{16\pi
}\int\limits_{\mathcal{M}}d^{N+1}x\sqrt{-g} \left( R-2\Lambda
-\left( \alpha \mathcal{F}\right) ^s \right) + \mathcal{I}_{\rm
matter},  \label{Action}
\end{equation}
where $R$ is scalar curvature, $\Lambda $ refers to the negative
cosmological constant (both $N+1$ dimensional), which is in general
equal to $-N(N-1)/2\chi^{2}$ for asymptotically AdS solutions, in
which $\chi$\ is a scale length factor, $\alpha $ is a constant and
$s$ is power of non-linearity and we choose $s=N/2$ \cite{Hendi1}.
We consider a null fluid as a matter field, whose action is
represented by $\mathcal{I}_{\rm matter}$ in Eq.~(\ref{Action}).

The field equations obtained by varying the action (\ref{Action})
with respect to the metric $g_{a b }$ and the gauge field $A_{a }$
read respectively
\begin{equation}
\mathcal{ G}_{a b} = R_{a b }-\frac{1}{2}R g_{a b } + \Lambda g_{a b
} = T_{a b}, \label{GravEq}
\end{equation}
and
\begin{equation}
\frac{1}{\sqrt{-g}}\partial _{a }\left( \sqrt{-g}F^{a b }\left(
\alpha F_{a b }F^{a b }\right) ^{N/2-1}\right) =0.  \label{MaxEq}
\end{equation}
The Maxwell tensor is $F_{ab} = \partial_a A_b -
\partial_b A_a$, with $\mathcal{F}=F_{a b }F^{a b }$ and $A_a$ is
vector potential. Here the energy-momentum (EMT) tensor is given by
\begin{equation}\label{emt}
T_{a b} =E_{a b } +T_{a b}^n
\end{equation}
where $T_{a b}^n = \zeta(v,r) n_{a}n_{b}$ with $\zeta(v,r)$ is the non-zero energy density of null fluid and
$n_a$ is a null vector such that
\begin{eqnarray}
n_{a} = \delta_a^0, n_{a}n^{a} = 0,
\end{eqnarray}
and the EMT  of the nonlinear electrodynamics is defined by
\begin{eqnarray}
E_{ab} =\alpha \left( \alpha \mathcal{F}\right) ^{N/2-1}\left(
\frac{1}{2}g_{a b }\mathcal{F-} NF_{a \lambda }F_{b }^{\;\lambda
}\right) , \label{emtM}
\end{eqnarray}
Now we consider ansatz for the spacetime which  expressed in terms
of Eddington advanced time coordinate (ingoing coordinate) $v$ is
given by
\begin{equation}
ds^{2} = -e^{\psi(v,r)}dv\left(f(v,r)e^{\psi(v,r)}dv + 2\epsilon
dr\right) + r^{2}(d \Omega_{N-1})^2, \label{Metric}
\end{equation}
where $(d \Omega_{N-1})^2 = \sum_{i=1}^{N-1} d\theta_i^2$, $v$ is a
null coordinate with
  $-\infty < v < \infty$, $r$ is a radial coordinate with
$0\leq r < \infty$. Here $\epsilon = \pm 1$. When $\epsilon = 1$,
the radial coordinate $r$ increases towards the future along a ray
$v = const$. When $\epsilon = -1$, the radial coordinate $r$
decreases towards the future along a ray $v =  const$.  In what
follows, we shall consider $\epsilon =1$. We wish to find the
general solution of the Einstein equation for the matter field given
by Eq.~(\ref{emt}) for the metric (\ref{Metric}), which contains two
arbitrary functions $\psi(v,r)$ and $f(v,r)$. It is the field
equation $G^0_1 = 0$ that leads to $ e^{\psi(v,r)} = g(v)$
\cite{sgad}. This could be absorbed by writing $d \tilde{v} = g(v)
dv$. Therefore the entire family of solutions we are searching for
is determined by a single function $f(v,r)$.

Henceforth, in this section, we adopt here a procedure similar to
Hendi \cite{Hendi1}, which we modify here to accommodate our
nonstatic case. Without loss of generality, we modify the vector
potential, in our case, as
\begin{equation}\label{pot}
A_{a} = h(v,r) \delta_a^v.
\end{equation}
In addition, we only consider purely radial electric \textit{ansatz
} doe the electromagnetic field which means only non vanishing
component of Maxwell tensor are given by
\begin{equation}\label{fvr}
F_{r v} = - F_{v r} =   \frac{\partial h}{\partial r},\;
\mbox{and}\; \mathcal{F}=-2\left( \frac{\partial h}{\partial
r}\right) ^{2},
\end{equation}
which is negative. The sign of the constant $\alpha $ will ensure
the real solutions. The power Maxwell invariant, $\left( \alpha
\mathcal{F}\right) ^{N/2}$, may be imaginary for positive $\alpha$,
when $N/2$ is fractional (for even dimension). Therefore, we set
$\alpha =-1$. For the vector potential (\ref{pot}), the non-linear
electromagnetic field equation (\ref{MaxEq}) leads to
\begin{equation}\label{meh}
r \frac{\partial^2 h}{\partial r^2} + \frac{\partial h}{\partial r}
= 0,
\end{equation}
which admits a solution $h(v,r) = q(v) \ln(r/\chi)$ and hence the
only non-vanishing component of  Maxwell tensor $F_{ab}$ takes the
form $F_{r v} = - F_{v r} = q(v)/r,$ where $q(v)$ is arbitrary. It
is interesting to note that the expression of the Maxwell tensor
$F_{ab}$ does not depend on the dimension and its value  is  similar
as $(2+1)$-dimensional BTZ solution \cite{BTZ1,BTZ2}. It is notable
that for higher dimensional linear Maxwell field equation, the
 Maxwell tensor depends on the dimensionality but in our case, the
 Maxwell tensor is proportional to $r^{-1}$ in all dimensions.
This results are consistent with the corresponding static case
\cite{Hendi1}.

The model considered is obtained from the EMT(\ref{emt}) which  is
such that the component $T_{vr}$ must be nonzero and $T^v_v=T^r_r$
for null energy condition.
 The nonzero components EMT would read as:
\begin{eqnarray}\label{emt2}
 T^r_v &=&\zeta(v,r),\;
  T_v^v = T_r^r
  =\frac{(N-1)}{2}2^{N/2}\left(\frac{q(v)}{r}\right)^N,
 \nonumber \\  T^{\theta_1}_{\theta_1}&=&T^{\theta_2}_{\theta_2} =\, .\, .\, .\, =
T^{\theta_{N-2}}_{\theta_{N-2}}=
\frac{2^{N/2}}{2}\left(\frac{q(v)}{r}\right)^N .
\end{eqnarray}
The part of the EMT, $T_{a b}^n $, can be considered as the
component of the matter field that moves along the null
hypersurfaces $v =$ constant. Type II fluid is the special case in
which the  EMT  has double null eigenvector. The only observed
occurrence of this form EMT is for zero rest-mass field when they
represent radiation \cite{he}.  The vector field specifies the
direction in which the radiation is moving.  The physical phenomena
which can be modeled by null fluid solutions can be a beam of
neutrinos which are assumed for simplicity to be massless. Whereas
the EMT of Type I fluid
 \begin{equation} T^a_b =
  \mbox{{{diag}}}[\rho,\; P_1,\; P_2,\; P_3,\; \ldots,\; P_N].
\label{emt3}
\end{equation}
The EMT of a Type I fluid has only one timelike eigenvector
\cite{ww,sgad,he}.  This is form of the EMT for all observed matter
fields with non-zero rest mass and also for the zero rest mass
except the Type II fluid discussed above \cite{he}.

 It is easy to show that the solutions for
$f(v,r)$ is obtained by solving the field equations (\ref{GravEq}).
The ($v,\;v$) component of the equation is integrated to give the
general solution as
\begin{equation}\label{sol}
f(v,r)=\frac{r^{2}}{\chi^{2}}-r^{2-N}\left[
{M(v)+}2^{N/2}q(v)^{N}\ln \left(\frac{r}{\chi}\right)\right] ,
\label{F(r)}
\end{equation}
where $M(v)$ and $q(v)$ are the  arbitrary functions which are
related to mass and charge parameters, respectively,  and which are
restricted only by the energy conditions. Thus we have $N$+1
dimensional nonstatic solution of Einstein field equations with the
metric (\ref{Metric}) for a charged null fluid defined by the
energy-momentum tensor (\ref{emt}), i.e., we have a kind of $N$+1
dimensional charged radiating metric.  Whereas when just $N=2$, it
reduces to the Vaidya-like black hole solution obtained by Husain
\cite{vh} in $2+1$ dimension. Hence, we refer our solution as
nonstatic BTZ like solution in $N$+1 dimensional Einstein-nonlinear
Maxwell gravity representing gravitational collapse of a charged
null fluid in an anti-de Sitter spacetime.  The metric function
$f(v,r)$, presented here, differ from the linear higher dimensional
Bonnor-Vaidya solutions \cite{bv,sgad}. It is notable that the
electric charge term in the linear case is proportional to
$r^{-2(N-2)}$, but in the presented metric function, nonlinear case,
this term is logarithmic as in the corresponding static
case\cite{Hendi1} As, in the case of static black hole
\cite{Hendi1}, the electric field of
 our nonstatic BTZ-like solutions is of similar nature as $(2+1)$-dimensional BTZ black holes. From the ($r,\;v$) component of
field equation and (\ref{sol}), then
\begin{equation}
\zeta(v,r)  = \frac{N-1}{2r^{N-1}} \left[ \dot{M}(v) + N
q(v)^{(N-1)} \dot{q}(v)\ln \left(\frac{r}{\chi}\right) \right].
\label{eq:mu}
\end{equation}
The special case in which $\dot{M}(v) = 0 $ and $\dot{q}(v) = 0$,
Eq.~(\ref{sol})  after the transformation
\begin{equation}
dt = dv - \left(\frac{r^{2}}{\chi^{2}}-r^{2-N}\left[
{M(v)+}2^{N/2}q(v)^{N}\ln \left(\frac{r}{\chi}\right) \right]
\right)^{-1} dr,
\end{equation}
 leads to static black hole found by Hendi \cite{Hendi1}. Further in
$2+1$-dimensional limit ($N=2$), these solutions reduce to the well
known BTZ solution \cite{BTZ2}.

The vacuum state, namely, what is to be regarded as empty space, is
obtained  by  letting $M(v)\rightarrow 0$ and $q(v)\rightarrow 0$.
\begin{equation}
ds^2 = - \frac{ r^2}{\chi^2}  dv^2 + 2 dv dr + r^{2}(d
\Omega_{N-1})^2.
 \label{ime}
\end{equation}
When $M(v)= �1$ and $q(v)=O$, one gets anti-de Sitter like
spacetime
\begin{equation}
ds^2 = - (1+\frac{  r^2}{\chi^2})  dv^2 + 2 dv dr + r^{2}(d
\Omega_{N-1})^2.
 \label{ime1}
\end{equation}

\subsection{Energy conditions and  horizons} The family of solutions discussed
here, in general, belongs to Type II fluid defined in \cite{he}.
When $M=q$= constant, the matter field degenerates to type I fluid
\cite{ww,sgad}. In the rest frame associated with the observer, the
energy-density of the matter will be given by
\begin{equation}
\mu = T^r_v,\hspace{.1in}\rho = - T^t_t = - T^r_r  \label{energy}
\end{equation}
 and the principal pressures are $P_i = T^i_i$ (no sum convention).

\noindent \emph{a) The weak energy conditions} (WEC): The energy
momentum tensor obeys inequality $T_{ab}w^a w^b \geq 0$ for any
time-like vector \cite{he}, i.e., $\zeta \geq 0,\hspace{0.1 in}\rho
\geq 0,\hspace{0.1 in} P   \geq 0,$. Both WEC and SEC, for a Type II
fluid, are identical \cite{ww,sgad}.
 Clearly, for WEC and SEC to be satisfied, we require that $\dot{M}(v) \geq 0$,
$\dot{q}(v) \geq 0$.  Physically this means that the matter within a
radius r increases with time, which corresponds to an implosion.

\noindent {\emph{b) The dominant energy conditions }}(DEC): For any
time-like vector $w_a$, $T^{ab}w_a w_b \geq 0$, and $T^{ab}w_a$ is
non-space-like vector, i.e., $ \zeta \geq 0,\hspace{0.1 in}\rho \geq
P \geq 0.$    It is easy to show that DEC holds as well.

In order to further discuss the physical nature of our solutions, we
introduce the kinematical parameters. Following York \cite{jy}, a
null-vector decomposition of the metric  is made of the form
\begin{equation}\label{gab}
g_{ab} = - n_a l_b - l_a n_b + \gamma_{ab},
\end{equation}
where,
\begin{eqnarray}
n_{a} = \delta_a^v, \: l_{a} = \frac{1}{2} f(v,r) \delta_{a}^v +
\delta_a^r, \label{nva}
 \\
\gamma_{ab} = r^2 \delta_a^{\theta_1} \delta_b^{\theta_1} + r^2
\left[\prod_{j=1}^{i-1} {\theta}_j^2  \right] \delta_a^{\theta_i}
\delta_b^{\theta_i}, \label{nvb}
\\
l_{a}l^{a} = n_{a}n^{a} = 0 \; ~l_a n^a = -1, \nonumber \\ l^a
\;\gamma_{ab} = 0; \gamma_{ab} \; n^{b} = 0, \label{nvd}
\end{eqnarray}
The expansion of the null rays parameterized by $v$ is given by
\begin{equation}\label{theta}
\Theta = \nabla_a l^a - K,
\end{equation}
where the $\nabla$ is the covariant derivative and the surface
gravity is
\begin{equation}\label{sggb}
K = - n^a l^b \nabla_b l_a.
\end{equation}  We now consider the evolution of
the apparent horizon (AH)   for the metric (\ref{Metric}). The AH is
the outermost marginally trapped surface for the outgoing photons.
The AH can be either null or space-like, that is, it can 'move'
causally or acausally \cite{jy}. It gives the equation
\begin{equation}\label{mass}
m(r, v; c_i ) = \frac{r}{2}\; \mbox{for}\; r_{AH}(v; c_i )  .
\end{equation}
The $c_i$ are any constants that appear in the mass function. The
long time $  v \rightarrow \infty $ limit of $r_{AH}$ gives the
asymptotic radius of the AH as a function of the constants $c_i$
This is a measure of the black hole mass for asymptotically flat or
(anti)-de Sitter spacetimes, namely
\begin{equation}\label{mass1}
M_{BH} := \lim_{v\rightarrow\infty }  r_{AH} (v; c)/2.
\end{equation}
The AHs are defined as surface such that $\Theta \simeq 0$ or
$f(v,r)=0$ \cite{jy}. Thus

\begin{equation}
r_{IAH} =\chi\exp \left\{ -\frac{1}{N}\mbox{LambertW}\left(0,y
\right) -\frac{M(v)}{ 2^{N/2}q(v)^{N}}\right\},
\end{equation}
\begin{equation}
r_{OAH} =\chi\exp \left\{ -\frac{1}{N}\mbox{LambertW}\left(-1,y
\right) -\frac{M(v)}{ 2^{N/2}q(v)^{N}}\right\},
\end{equation}
where \[y= \left( \frac{-N\chi^{N-2}e^{\left(
\frac{-NM(v)}{2^{N/2}q(v)^{N}}\right)}}{ 2^{N/2}q(v)^{N}} \right),\]
 and the LambertW function satisfies
 \[\mbox{LambertW}(x)\exp \left[\mbox{LambertW}(x)\right]=x. \] The pure charged case ($M(v)$ =
0) is also important, since then
\begin{equation}\label{solpc}
f(v,r) = -{\frac {{r}^{2}}{{\chi}^{2}}}+{r}^{(2-N)}{2}^{N/2}  q
\left( v \right)^{N}\ln  \left( {\frac {r}{\chi}} \right)
\end{equation}
and we have horizons without mass
\begin{equation}\label{ahi}
r_{IAH} = \chi \exp \left[-\frac{1}{N}\mbox{LambertW}\left(0,-
\frac{N \chi^{(N-2)}}{q(v)^N 2^{(N/2)}}\right) \right].
\end{equation}
\begin{equation}\label{aho}
r_{OAH} = \chi \exp \left[-\frac{1}{N}\mbox{LambertW}\left(-1,-
\frac{N \chi^{(N-2)}}{q(v)^N 2^{(N/2)}}\right) \right].
\end{equation}
The uncharged case ($q(v)\rightarrow 0$) is interesting, since then
and apparent horizon is $r_{AH}=\exp({\frac {\ln  \left( M \left( v
\right) {\chi}^{2} \right)}{N}})$. It is clear that presence of the
nonlinear Maxwell source produces a change in the location of the
AHs. Such a change could have a significant effect in the dynamical
evolution of the black hole horizon.  The timelike limit surface
(TLS) for a black hole, with a small luminosity, is locus where
$g_{vv} = 0$ \cite{jy}. Here one sees that $\Theta = 0$, implies
$f(v,r_{AH})=0$ or $g_{vv}(r=r_{AH}) = 0$ implies that $r=r_{AH}$ is
TLS and thus in our case AH and TLS coincide.

\section{Causal Structure of Singularities}
The easiest way to detect a singularity in a space-time is to
observe the divergence of some invariants of the Riemann tensor. The
Kretschmann scalar ($K = R_{abcd} R^{abcd}$, $R_{abcd}$ is the $N+1$
Riemann tensor) for the metric (\ref{Metric}) reduces to
\begin{equation}
\mbox{K} = \left(\frac{\partial^2 f}{\partial
r^2}\right)^2+\frac{2(N-1)}{r^4}\left(\frac{\partial f}{\partial
r}\right)^2 + \frac{2(N-2)(N-1))}{r^4} (f)^2,\label{ks}
\end{equation}
which diverges and so is energy density (\ref{eq:mu}) at $r=0$ for
$f\neq0$ and $N\geq 2$ indicating presence of scalar polynomial
singularity. The physical situation is that of a radial influx of
charged null fluid in the region of the anti-de Sitter universe. The
first shell arrives at $r=0$ at time $v=0$ and the final at $v=T$. A
central singularity of growing mass developed at $r=0$.
  For $ v < 0$ we have $M(v)\;=\;q(v)\;=\;0$, i.e., $N+1$ dimensional vacuum metric (\ref{ime}),
and for $ v > T$, $\dot{M}(v)\;=\;\dot{q}(v)\;=\;0$, $M(v)$ and
$q(v) $ are positive definite.  The metric for $v=0$ to $v=T$ is
$N$+1 dimensional nonstatic BTZ-like  radiating solution derived
above, and for $v>T$ we have the $N$+1 dimensional static BTZ-like
solution \cite{hendi}
\begin{eqnarray}
ds^2 & = &\left(- \frac{r^{2}}{\chi^{2}}-r^{2-N}\left[ {M+}2^{N/
2}q^{N}\ln \left(\frac{r}{\chi}\right)\right]\right)  dv^2 + 2 dv dr
\nonumber \\ & & + r^{2}(d \Omega_{N-1})^2.\label{eq:ome}
\end{eqnarray}
 Radial ($ \theta$ and $ \phi \,=\,const$.) null geodesics
of the metric (11) must satisfy the null condition
\begin{equation}
\frac{dv}{dr} = \frac{2}{\left[ \frac{r^{2}}{\chi^{2}}-r^{2-N}\left(
{M(v)}+2^{N/2}q(v)^{N}\ln \left(\frac{r}{\chi}\right)\right)
\right]} \label{eq:de1}
\end{equation}
Clearly, the above differential equation has a singularity at $r=0$,
$v=0$. The nature (a naked singularity or a black hole) of the
collapsing solutions can be characterized by the existence of radial
null geodesics coming out from the singularity.  The nature of the
singularity can be analyzed by techniques in \cite{psj}. To proceed
further, we choose
\begin{equation}
M(v) = \lambda v^{(N-2)} \; (\lambda > 0) \;\mbox{and}\;
2^{N/2}q(v)^N = \mu^2 v^{(N-2)} (\mu
> 0), \label{eq:mv}
\end{equation}
for $0 \leq v \leq T$ \cite{psj,lz}. Let $Y \equiv v/r$ be the
tangent to a possible outgoing geodesic from the singularity.

 In order to determine the nature of the
limiting value of $Y$ at $r=0$, $v=0$ on a singular geodesic, we let
$ Y_{0} = \lim_{r \rightarrow 0 \; v\rightarrow 0} Y =
\lim_{r\rightarrow 0 \; v\rightarrow 0} \frac{v}{r} $.
 Using
(\ref{eq:de1})   and L'H\^{o}pital's rule we get
\begin{eqnarray}
Y_{0} & = & \lim_{r\rightarrow 0 \; v\rightarrow 0} Y =
\lim_{r\rightarrow 0 \; v\rightarrow 0} \frac{v}{r}=
 \lim_{r\rightarrow 0 \; v\rightarrow 0} \frac{dv}{dr}  \nonumber  \\
& = &  \lim_{r\rightarrow 0 \; v\rightarrow 0}
\frac{2}{\frac{r^2}{\chi^2} - \left(\lambda  Y^{N-2} + \mu^2 Y^{N-2}
\ln \left(\frac{r}{\chi}\right)\right)} \label{eq:ae}
\end{eqnarray}

This is the equation which would ultimately decide the end state of
collapse:
 a  black hole or a  naked singularity. Thus by analyzing
this algebraic equation, the nature of the singularity can be
determined.  The central shell focusing singularity would atleast be
locally naked (for brevity we have addressed it as naked throughout
this paper), if Eq.~(\ref{eq:ae})
 admits one or more positive real roots \cite{psj}.
 The values of the roots give the tangents of the escaping geodesics
near the singularity. When there are no positive real roots to
Eq.~(\ref{eq:ae}), there are no out going future directed null
geodesics emanating from the singularity. Thus, the occurrence of
positive roots would imply the violation of the strong CCC, though
not necessarily of the weak form. Hence in the absence of positive
real roots, the collapse will always lead to a black hole. Clearly,
Eq.~(\ref{eq:ae}) do not admit positive roots \cite{psj} and no
radial future null geodesics terminate at the singularity.  Thus
referring to the above discussion, the collapse proceed to form a
black hole.

\subsection{CLOSING REMARKS}
In this paper we have demonstrated a construction $N+1$ dimension
charged BTZ-like nonstatic black holeb solutions, namely $N+1$
dimensional Vaidya-form of BTZ black holes.  We have obtained the
general black hole solutions for charged null fluid for the metric
\textit{ansatz }~(\ref{Metric}) in the Eddington advanced time
coordinate of $N+1$ dimensional gravity coupled with a nonlinear
electrodynamics theory as a power $s$ of the Maxwell invariant. This
yields, in $2+1$ dimension and in static limit, the  BTZ solutions
\cite{BTZ2}. We have used the solution to discuss the consequence of
nonlinear electrodynamics on the structure and location of the
horizons these radiating black holes. The AHs are obtained exactly
by method developed by York \cite{jy} and it is also shown that the
AHs of these black holes coincides with TLS as it should be.

The gravitational collapse of spherical matter in the form of null
fluid described by Vaidya metric \cite{pc} is well studied for
investigating CCC. It clear that the central shell focusing
singularity can be naked or covered depending upon on the choice of
initial data.  The scenarios considered so far (see \cite{psj} for
details) are spherically symmetric and asymptotically flat. We may
then ask if the occurrence of a naked singularity in these cases is
an artefact of the special symmetry. In the presence of negative
cosmological term and departure from spherical symmetry one can
expect the occurrence of major changes. When a negative cosmological
constant is introduced, the spacetime will become asymptotically
anti-de Sitter spacetime.  The present case is an example of non
spherical symmetry as well as non asymptotic flatness, and we have
shown both uncharged and charged collapse of null fluid does not
yield naked singularities, but always black holes.  This shows that
non-spherical collapse of null fluid (charged or uncharged)
discussed here in a negative cosmological background does nor
violate CCC and the result is independent of spacetime dimension. We
therefore conclude that CCC is actually respected for our nonstatic
BTZ-like solution in all dimensions

In conclusion, we have derived a new class of $N+1$ dimensional
charged null fluid collapse solutions  of the Einstein equations
with a negative cosmological constant. As a special case these
solutions can be reduced to the BTZ black hole.  The study of
geometric properties, causal structures and thermodynamics of these
black hole solutions will be subject of the future project. As final
remark it would be interesting to explore  the extensions of the
solutions presented here in more general context, e.g., to see how
the results get modified with the inclusion of  Gauss-Bonnet
combination of quadratic curvature terms and, in general, for the
Lovelock polynomial in the action (\ref{Action}), which is left for
future investigation \cite{sgg}.
\begin{acknowledgements}
The work is supported by  university grant commission (UGC) major
research project grant F. NO. 39-459/2010 (SR).
\end{acknowledgements}

\end{document}